\documentclass[aps,prb,preprint,groupedaddress,citeautoscript,nopacs]{revtex4}

\usepackage{graphicx}
\usepackage{multirow}
\usepackage{color}
\usepackage{bm}
\usepackage{ulem}
\usepackage{times}
\usepackage{amsmath,bm,amsfonts}
\usepackage{dcolumn}
\usepackage{graphicx}
\usepackage{latexsym}
\usepackage{natbib}

\begin{document}

\title{Two-dimensional higher-order topology in monolayer graphdiyne}

\author{Eunwoo \surname{Lee}}\thanks{equal contribution}
\affiliation{Department of Physics and Astronomy, Seoul National University, Seoul 08826, Korea}

\affiliation{Center for Correlated Electron Systems, Institute for Basic Science (IBS), Seoul 08826, Korea}

\affiliation{Center for Theoretical Physics (CTP), Seoul National University, Seoul 08826, Korea}

\author{Rokyeon \surname{Kim}}\thanks{equal contribution}
\affiliation{Department of Physics and Astronomy, Seoul National University, Seoul 08826, Korea}

\affiliation{Center for Correlated Electron Systems, Institute for Basic Science (IBS), Seoul 08826, Korea}

\affiliation{Korea Institute for Advanced Study (KIAS), Seoul 02455, Korea}

\author{Junyeong \surname{Ahn}}
\affiliation{Department of Physics and Astronomy, Seoul National University, Seoul 08826, Korea}

\affiliation{Center for Correlated Electron Systems, Institute for Basic Science (IBS), Seoul 08826, Korea}

\affiliation{Center for Theoretical Physics (CTP), Seoul National University, Seoul 08826, Korea}

\author{Bohm-Jung \surname{Yang}}
\email{bjyang@snu.ac.kr}
\affiliation{Department of Physics and Astronomy, Seoul National University, Seoul 08826, Korea}

\affiliation{Center for Correlated Electron Systems, Institute for Basic Science (IBS), Seoul 08826, Korea}

\affiliation{Center for Theoretical Physics (CTP), Seoul National University, Seoul 08826, Korea}

\date{\today}

\begin{abstract}
Based on first-principles calculations and tight-binding model analysis, we propose monolayer graphdiyne as a candidate material for a two-dimensional higher-order topological insulator protected by inversion symmetry. Despite the absence of chiral symmetry, the higher-order topology of monolayer graphdiyne is manifested in the filling anomaly and charge accumulation at two corners. Although its low energy band structure can be properly described by the tight-binding Hamiltonian constructed by using only the $p_z$ orbital of each atom, the corresponding bulk band topology is trivial. The nontrivial bulk topology can be correctly captured only when the contribution from the core levels derived from $p_{x,y}$ and $s$ orbitals are included, which is further confirmed by the Wilson loop calculations. We also show that the higher-order band topology of a monolayer graphdyine gives rise to the nontrivial band topology of the corresponding three-dimensional material, ABC-stacked graphdiyne, which hosts monopole nodal lines and hinge states.
\end{abstract}


\maketitle

\section{Introduction}
Bulk-boundary correspondence is a fundamental property of topological phases. In conventional topological insulators, the gapped bulk states in $d$-dimensions support metallic states in $(d-1)$-dimensional surfaces. Such a conventional bulk-boundary correspondence, however, is violated in a class of topological crystalline insulators, so-called higher-order topological insulators (HOTIs). The gapless excitations of a HOTI in $d$-dimensions are localized, instead, in a subspace with a dimension lower than $(d-1)$, such as corners or hinges, when the global shape of the material preserves the crystalline symmetry relevant to the nontrivial band topology~\cite{schindler2018higher2,schindler2018higher,wang2019higher,yue2019symmetry,wieder2018axion,ahn2018higher,benalcazar2017quantized,langbehn2017reflection,geier2018second,khalaf2018higher,benalcazar2019quantization}. 
Recently, rhombohedral bismuth has been identified as the first example of 3D HOTIs hosting helical hinge states~\cite{schindler2018higher2}. Also there are other candidate materials of 3D HOTIs including SnTe with strain along the (100) direction~\cite{schindler2018higher}, transition metal dichalcogenides MoTe$_2$ and WTe$_2$, hosting helical hinge states~\cite{wang2019higher}, and also Bi$_{2-x}$Sm$_x$Se$_3$ with chiral hinge states~\cite{yue2019symmetry}.  

In 2D, one can also consider second-order topological insulators (SOTIs) hosting localized corner states~\cite{benalcazar2017quantized,langbehn2017reflection,khalaf2018higher,geier2018second,benalcazar2019quantization,song2017d}. Contrary to 3D HOTIs, however, there are only few theoretical proposals for candidate materials realizing 2D SOTIs~\cite{ahn2019failure,khalaf2018higher,benalcazar2019quantization,lee2019fractional,geier2018second}.
For instance, phosphorene is one candidate for 2D HOTIs proposed based on simple tight-binding model calculations~\cite{ezawa2018minimal}. However, in this system, the corner states originate from the charge polarization~\cite{ezawa2018minimal}, which is a 1D topological invariant. Hence, in a strict sense, phosphorene is not a genuine 2D HOTI.  Another candidate is twisted bilayer graphene, which can also exhibit higher-order band topology at small twist angles. However, to realize the corner states, the twist angle as well as the chemical potential should be controlled simultaneously~\cite{ahn2019failure}. 

In general, among the higher-order TIs, only $d$-dimensional $k$th-order TIs with $1<k<d$ have stable band topology with Wannier obstruction while $d$-dimensional $d$th-order TIs with zero-dimensional corner charges are obstructed atomic insulators. Thus, in 3D, a SOTI has Wannier obstruction while a third-order TI (TOTI) is atomic. On the other hand, in 2D, HOTIs are always obstructed atomic insulators. In $d$-dimensional $d$th-order TIs, additional chiral or particle-hole symmetry is required to protect corner states which appear as in-gap states separated from bulk states~\cite{khalaf2018higher,geier2018second,benalcazar2019quantization,langbehn2017reflection}. Since chiral or particle-hole symmetry usually does not exist in insulators, realizing 2D SOTIs in materials is generally considered difficult. 

However, even if the chiral symmetry of 2D SOTIs is broken, the material still inherits the nontrivial band topology of 2D SOTIs. The resulting state is called a 2D TOTI~\cite{cualuguaru2019higher,okuma2019topological,hwang2019fragile}. The key idea is that even if the in-gap states are merged into the bulk states, the nontrivial band topology manifests in the {\it filling anomaly}~\cite{wieder2018axion, benalcazar2019quantization,song2017d}. Namely, the half-filling condition can never be satisfied as long as the symmetry associated with the second-order band topology is preserved. The half-filling condition can be satisfied only when extra electrons or holes are added, which naturally leads to charge accumulation or depletion at corners. Therefore, the hallmark of 2D HOTIs lacking chiral symmetry (or simply 2D TOTIs), is the existence of filling anomaly at half-filling and the accumulation or depletion of a half-integral charge at corners after adding extra electrons or holes~\cite{cualuguaru2019higher,okuma2019topological,hwang2019fragile}.

In this work, we theoretically propose monolayer graphdiyne (MGD) as the first realistic candidate material for 2D HOTIs. Namely, MGD is a 2D SOTI when chiral symmetry exists, but in real systems, it appears as a 2D TOTI due to the lack of chiral symmetry. Based on first-principles calculations and tight-binding model analysis, we show that a MGD is a HOTI characterized by a 2D topological invariant $w_2$, that is quantized when the system is invariant under inversion $P$ symmetry. Although corner states are not guaranteed to be in the gap due to the lack of chiral symmetry in MGD, the nontrivial higher-order band topology is manifested in filling anomaly and the corner charge accumulation. Also, it is found that although the low energy excitation can be well-described by using the state derived from the $p_z$ orbitals of each atom, the relevant band topology of such a simplified model is trivial. To describe the nontrivial band topology, it is essential to include the contribution from the core levels derived from $p_{x,y}$ and $s$ orbitals. Finally, it is shown that the non-trivial $w_2$ of MGD gives rise to the topological band structure of the corresponding three-dimensional material, ABC stacked graphdiyne, hosting nodal lines with $Z_2$ monopole charges~\cite{nomura2018three,ahn2018band}. Due to the nonzero $w_2$ in a range of 2D momentum subspaces with fixed vertical momentum $k_z$, the ABC stacked graphdiyne host hinge states for the corresponding range of $k_z$.

\section{results}
\subsection{Topological invariant of monolayer graphdiyne}
\begin{figure}[t]
\centering
\includegraphics[width=8.5 cm]{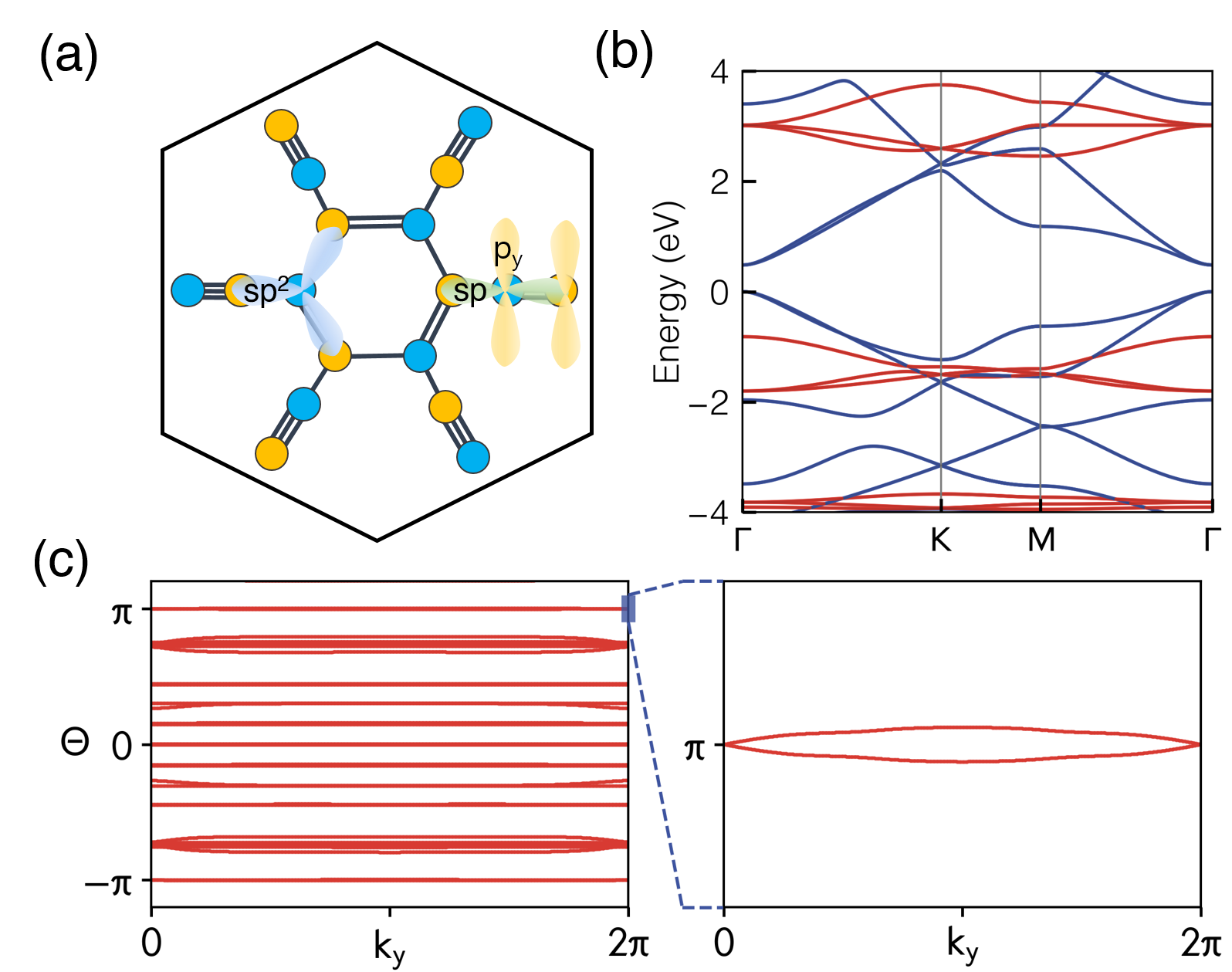}
\caption{
Lattice structure, band structure, and the Wilson loop spectrum.
(a) A schematic figure describing the unit cell and the relevant atomic orbitals of a monolayer graphdiyne (MGD) composed of 18 carbon atoms. Two sublattices are marked with yellow and blue colors, respectively. Since MGD has a bipartite lattice structure, chiral symmetry exists when only the nearest neighbor hoppings are considered. (b) The band structure of MGD obtained by first-principles calculations, which shows approximate chiral symmetry near the Fermi level. Since $p_z$ orbitals are odd while $s$, $p_x$, $p_y$ orbitals are even under the mirror $M_z:z\rightarrow-z$ operation, the energy spectrum from $p_z$ orbitals (blue) is not hybridized with that from $s$, $p_x$, $p_y$ orbitals (red). (c) The Wilson loop spectrum computed including $s$, $p_x$, $p_y$, $p_z$ orbitals. The spectrum exhibits a crossing point at $k_y=0$, $\theta=\pi$, which indicates $w_2=1$.
}
\label{fig:graphdiynesublattice}
\end{figure}

MGD is an experimentally realized planar carbon system composed of $sp^2$-$sp$ carbon network of benzene rings connected by ethynyl chains~\cite{haley1997carbon,li2010architecture}. Fig.~\ref{fig:graphdiynesublattice}(a) describes the unit cell of MGD composed of 18 carbon atoms. Since spin-orbit coupling (SOC) is negligible, MGD can be regarded as a spinless fermion system. The lattice has inversion $P$, time-reversal $T$, a two-fold rotation about the $x$-axis $C_{2x}$, a six-fold rotation about the $z$-axis $C_{6z}$, and a mirror $M_z:z\rightarrow-z$ symmetries. 
Also, due to the bipartite lattice structure, MGD has chiral (or sublattice) symmetry when only the nearest neighbor hopping between different sublattices is considered, similar to the case of graphene. 

Fig.~\ref{fig:graphdiynesublattice}(b) shows the band structure along high symmetry directions obtained by first-principles calculations. Here the blue color indicates the bands derived from $p_z$ orbitals while the red color denotes the bands derived from $s$, $p_x$, $p_y$ orbitals. Since $p_z$ orbitals are odd whereas $s$, $p_x$, $p_y$ orbitals are even under $M_z$ symmetry, the energy spectrum from $p_z$ orbitals is not hybridized with that from $s$, $p_x$, $p_y$ orbitals. One can see that the low-energy band structure has approximate chiral symmetry and can be effectively described by using only $p_z$ orbitals. However, to capture the higher-order band topology, the core electronic states derived from $s$, $p_x$, $p_y$ orbitals should be included as shown below.

In general, a 2D $P$-symmetric spinless fermion system carries three $Z_2$ topological invariants, $w_{1x}$, $w_{1y}$, and $w_2$~\cite{ahn2018band}. In terms of the sewing matrix $G_{mn}(\textbf{k})=\langle u_{m\textbf{-k}}|P|u_{n\textbf{k}}\rangle$, $w_{1a}$ ($a=x,~y$) is given by its 1D winding number as
\begin{align}
w_{1a}= -\frac{i}{2}\oint_{C}\nabla_{\textbf{k}_a}\log\det G(\textbf{k})\mod 2,
\end{align}
which is equivalent to the quantized Berry phase $\Phi_a$, in a way that $\Phi_a=\pi w_{1a}$ (mod $2\pi$)~\cite{zak1989berry,xiao2010berry}.
$w_{1a}$ can be determined by using relation $(-1)^{w_{1a}} = \prod_{i=1}^{2}\xi(\Gamma_{ia})$ where $\xi(\Gamma_{ia})$ ($i=1,~2$) is the product of parity eigenvalues of occupied states at the time-reversal invariant momentum (TRIM) $\Gamma_{ia}$, and $\Gamma_{1a}$, $\Gamma_{2a}$ are two TRIMs along the reciprocal lattice vector $G_a$. Since $w_{1a}$ is equivalent to the quantized charge polarization along the $G_a$ direction, it can be considered as a 1D topological invariant.

On the other hand, $w_2$ is a 2D topological invariant given by the 2D winding number of the sewing matrix $G(\textbf{k})$ (modulo 2)~\cite{ahn2018higher}, and measures the higher-order band topology of $P$-symmetric spinless fermion systems~\cite{ahn2018higher,ahn2019failure,wieder2018axion}. $w_2$ can be determined by 
\begin{align}
    (-1)^{w_2}=\prod_{i=1}^{4}(-1)^{[N_{\text{occ}}^{-}(\Gamma_i)/2]},
\end{align}
where $N_{\text{occ}}^{-}(\Gamma_i)$ is the number of occupied bands with odd parity at the TRIM $\Gamma_i$ and the square bracket $[\alpha]$ indicates the greatest integer value of $\alpha$~\cite{song2018diagnosis,fu2007topological,kim2015dirac,hughes2011inversion,turner2012quantized,po2017symmetry}.

Using the band structure from first-principles calculations and the corresponding parity eigenvalues at TRIMs, one can easily show that MGD is a higher-order topological insulator with $w_{1x}=w_{1y}=0$ and $w_2=1$. One interesting feature of the MGD band structure is that we get $w_2=0$ when only the bands derived from $p_z$ orbitals are considered. Namely, although the low-energy band structure itself can be well-described by using only $p_z$ orbitals, the higher-order topology can be correctly captured only when the core electronic states derived from $s$, $p_x$, $p_y$ orbitals are included. This result can be confirmed by tight-binding analysis as well as first-principles calculations.
Since the bands derived from $p_z$ orbitals do not hybridize with those derived from other orbitals because of $M_z$ symmetry, one can compute the relevant topological invariants separately. 

The orbital dependence of $w_2$ can also be confirmed by Wilson loop calculations~\cite{ahn2018band,ahn2019failure,zhao2017p,fang2015topological}.
The Wilson loop is defined as a path ordered product of the exponential of Berry connections~\cite{yu2011equivalent,alexandradinata2014wilson},
\begin{align}
W_{(k_1+2\pi,k_2)\leftarrow(k_1,k_2)}&= \lim_{N\rightarrow\infty}F_{N-1}F_{N-2}\dotsb F_{1}F_{0},
\end{align}
where $[F_i]_{mn}=\langle u_m(\frac{2\pi(i+1)}{N},k_2)|u_n(\frac{2\pi i}{N},k_2)\rangle$, and its spectrum is gauge-invariant~\cite{yu2011equivalent}.
Computed along the $k_1$ direction parallel to the reciprocal lattice vector $\bf{G_1}$ from ($k_1, k_2$), the set of Wilson loop eigenvalues $\{e^{i\nu_i(k_2)} \}$ indicates the position of Wannier centers at given $k_2$, and the corresponding total charge polarization is given by $p_1(k_2)=\frac{1}{2\pi}\sum_{i=1}^{N_{occ}}\nu_i(k_2)$.
The Wilson loop spectrum of MGD shows that the charge polarization is zero in both $k_1$ and $k_2$ directions, which is consistent with the parity eigenvalue analysis (Fig.~\ref{fig:graphdiynesublattice}(c)). Also, the single linear-crossing at $(k_y,\nu)=(0,\pi)$ indicates $w_2=1$~\cite{ahn2018band}.
Projecting the Wilson loop operator into the $p_z$ ($s$, $p_x$, and $p_y$) orbital basis, we observe $w_2=0$ ($w_2=1$), which confirms that the higher-order band topology originates from the core electronic levels. The orbital dependence of $w_2$ can be further confirmed by computing the Wannier center of each orbital explicitly and also by computing the nested Wilson loop that measures the higher-order band topology~\cite{wieder2018axion,benalcazar2017quantized}.
In the presence of $PT$ symmetry, the Wilson loop spectrum can be divided into two sub-bands that are separated by a gap, each centered at $\nu=0$ or $\nu=\pi$, respectively. When the Wilson loop operator at given $k_2$ is diagonalized as 
\begin{align}
    W_{(k_1+2\pi,k_2)\leftarrow(k_1,k_2)}=\sum_{i}|\nu_i(k_2)\rangle e^{i \nu_i(k_2)}\langle \nu_i(k_2)|,
\end{align}
then, using the set of eigenvectors corresponding to the sub-band centered at $\nu=\pi$, the nested Wilson loop operator $\tilde{W}$ along the $\bf{G_2}$ vector is defined as
\begin{align}
\tilde{W}_{(k_2+2\pi)\leftarrow(k_2)}= \lim_{N\rightarrow\infty}\tilde{F}_{N-1}\tilde{F}_{N-2}\dotsb \tilde{F}_{0},
\end{align}
where $[\tilde{F}_i]_{mn}=\langle \nu_m(\frac{2\pi(i+1)}{N})|\nu_n(\frac{2\pi i}{N})\rangle$.
Since the determinant of the nested Wilson loop operator $\det(\tilde{W})$ is identical to $(-1)^{w_2}$~\cite{ahn2018band}, we find that $\det(\tilde{W})$ is nontrivial. Projecting the Wilson loop operator into the $p_z$ orbital basis, we observe that the corresponding nested Wilson loop spectrum is trivial. On the other hand, within the $s$, $p_x$, and $p_y$ orbital basis, the nested Wilson loop spectrum is nontrivial, which confirms that the higher-order band topology originates from the core electronic levels.

\subsection{Filling anomaly and corner charges}
The higher-order band topology of MGD with $w_2=1$ induces a pair of anomalous corner states~\cite{wang2019higher}. To see this, we consider a finite-size structure invariant under $P$ symmetry. When the system is chiral symmetric, there are two robust zero-energy in-gap states related by $P$ whose locations are fixed at $M_y$-invariant corners.
One way to see this is to use the topological classification proposed by Geier et al.~\cite{geier2018second}.
When a 2D system has commuting mirror and chiral symmetries, it belongs to BDI$^{M_{++}}$ class that supports nontrivial topology, and the corresponding insulator has zero modes at mirror invariant corners. In MGD, the $M_{y}$ symmetry commutes with chiral symmetry since $M_{y}$ symmetry does not mix different sublattices. On the other hand, when chiral symmetry and $M_x$ symmetry anti-commute, the system belongs to BDI$^{M_{+-}}$ class, which is topologically trivial, so there are no protected corner charges at $M_x$ invariant corners.

To be specific, we can find the location of corner states by studying the low-energy Hamiltonian of MGD near the $\Gamma$ point:
\begin{align}
H(k_x,k_y)&=\hbar v(k_x\Gamma_1+k_y\Gamma_2)+\Delta\Gamma_3,
\end{align}
where $\Gamma_1=\sigma_x\tau_x, \Gamma_2=\sigma_y\tau_x, \Gamma_3=\tau_z,$ and $\Gamma_4=\sigma_z\tau_x$. Corresponding symmetry representations are
$T=\sigma_x\tau_zK$, $P=\tau_z$, $M_x=\sigma_x\tau_z$, $M_y=\sigma_x$, and chiral symmetry $S$ is described by $S=\Gamma_{5}=\tau_y$.
Considering a disk-shaped geometry, the low-energy Hamiltonian in real space becomes $H=-i\tilde{\Gamma}_1(\theta)\partial_r-i\tilde{\Gamma}_2(\theta)r^{-1}\partial_{\theta}+\Delta(r)\Gamma_3$, where $\tilde{\Gamma}_1(\theta)=\cos\theta\Gamma_1+\sin\theta\Gamma_2$ and $\tilde{\Gamma}_2(\theta)=-\sin\theta\Gamma_1+\cos\theta\Gamma_2$. We assume that $\Delta(r)>0$ inside the bulk and $\Delta(r)<0$ outside the bulk. The edge Hamiltonian is obtained by applying the projection operator $P(\theta)=\frac{1}{2}(1+i\tilde{\Gamma}_1(\theta)\Gamma_3)$, leading to
$H_{\text{edge}}=-iR^{-1}\tilde{\sigma}_z\partial_{\theta}$ where $\tilde{\sigma}_z=P\tilde{\Gamma}_2(\theta)P^{-1}$.
On the edge, position-dependent mass terms that open the edge gap are allowed. Mass terms that are commuting with $P(\theta)$ and anti-commuting with $\tilde{\Gamma}_2(\theta)$ and $S$ are
$H_m=m_4(R,\theta)\Gamma_4+im_{25}(R,\theta)\tilde{\Gamma}_{2}\Gamma_{5}$. The corresponding edge Hamiltonian is
$H_{\text{edge}}=-iR^{-1}\tilde{\sigma}_z\partial_{\theta}+\tilde{m}_1(R,\theta)\tilde{\sigma}_x$, where $\tilde{m}_1=m_4+m_{25}$, $\tilde{\sigma}_x=P\Gamma_4P^{-1}=Pi\tilde{\Gamma}_{2}\Gamma_{5}P^{-1}$.
Here $PT$ symmetry requires $\tilde{m}_1(\theta+\pi)=-\tilde{m}_1(\theta)$ while
$M_y$ requires $\tilde{m}_1(x,y)=-\tilde{m}_1(x,-y)$.
Thus, the mass gap is closed at $M_y$-invariant corners leading to localized corner charges.

 Due to the degeneracy of the two in-gap states, the half-filling condition cannot be satisfied as long as $P$ is preserved, which is known as the filling anomaly~\cite{benalcazar2019quantization,wieder2018axion,song2017d}. When chiral symmetry is broken, the two degenerate in-gap states can be shifted from zero energy and even merged to the valence (conduction) bands generating a single hole (electron) at half-filling. Namely, the presence of an odd number of holes (electrons) in the valence (conduction) band is the manifestation of the higher-order band topology of generic 2D HOTIs lacking chiral symmetry~\cite{wieder2018axion,cualuguaru2019higher,okuma2019topological,hwang2019fragile}. To resolve the filling anomaly while keeping $P$, one needs to add an extra electron (or hole), which leads to the accumulation or depletion of a half-integral charge at each $M_y$-invariant corner.

Explicitly, let us illustrate how the number of states can be counted in a finite-size HOTI with $w_2=1$ following Ref.~\onlinecite{wieder2018axion}. When the total number of states is $N$ and chiral symmetry exists, there are $N/2-1$ occupied and unoccupied states, respectively, since two in-gap states appear at zero-energy. When chiral symmetry is broken while inversion is preserved, the two in-gap states related by inversion are absorbed into either the valence or the conduction bands simultaneously. Then the number of states below the gap is $n_{\text{gap}}=N/2\pm1$. 
Therefore $n_{\text{gap}}=(\frac{N}{2}+$ an odd integer) when $w_2=1$, whereas $n_{\text{gap}}=(\frac{N}{2}+$ an even integer) when $w_2=0$.
When the valence (conduction) band of a system with $w_2=1$ is completely occupied (unoccupied) by adding electrons or holes additionally, a half-integral electric charge should be accumulated (or depleted) at two corners related by inversion.

\begin{figure}[t]
\centering
\includegraphics[width=8.5 cm]{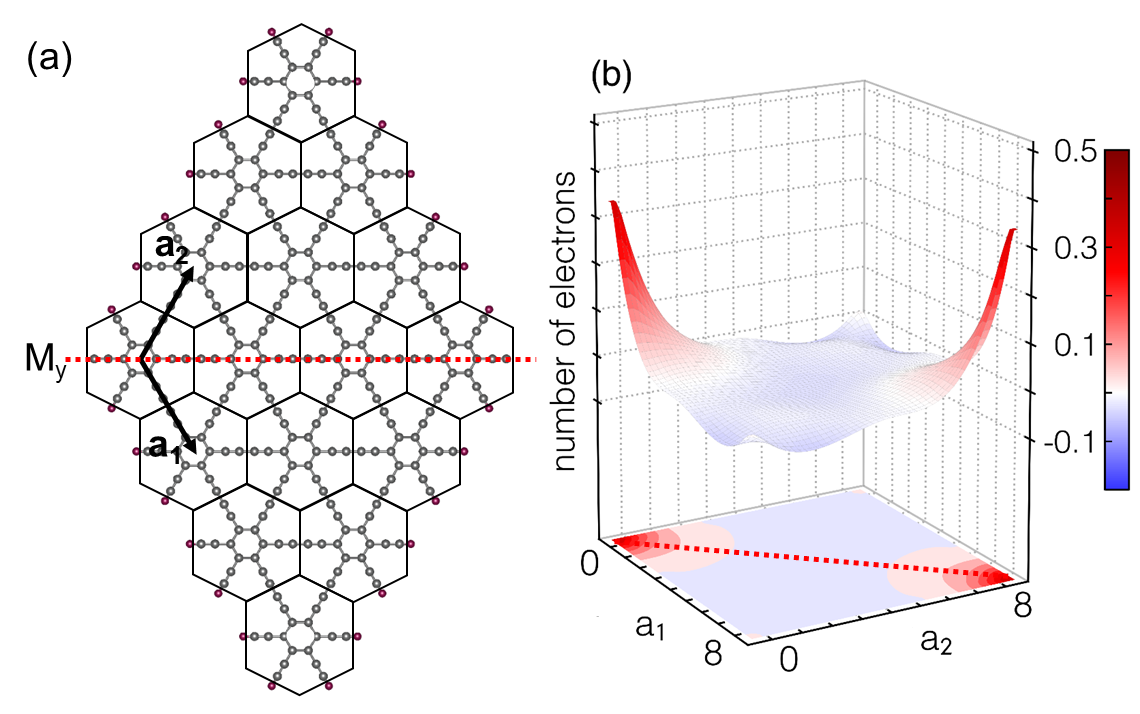}
\caption{
Geometry of a finite-size structure and corner charges.
(a) A finite-size structure of MGD preserving $P$ and $M_y$ symmetries designed to observe the higher-order band topology protected by $P$ symmetry. The red horizontal line indicates the $M_y$-invariant line. (b) Electron density distribution for a finite-size MGD composed of $9\times9$ unit cells where hydrogen atoms are attached at every corner except at the two $M_y$-invariant corners. To resolve the filling anomaly, we add one electron to MGD additionally and fill the valence band completely. The accumulated or depleted charges with respect to the half-filled configuration are plotted. Here a half-integral charge accumulated at each $M_y$-invariant corner appears due to the nontrivial 2D topological invariant $w_2=1$. 
} \label{fig:finitesize}
\end{figure}

\begin{figure*}[t]
\centering
\includegraphics[width=15 cm]{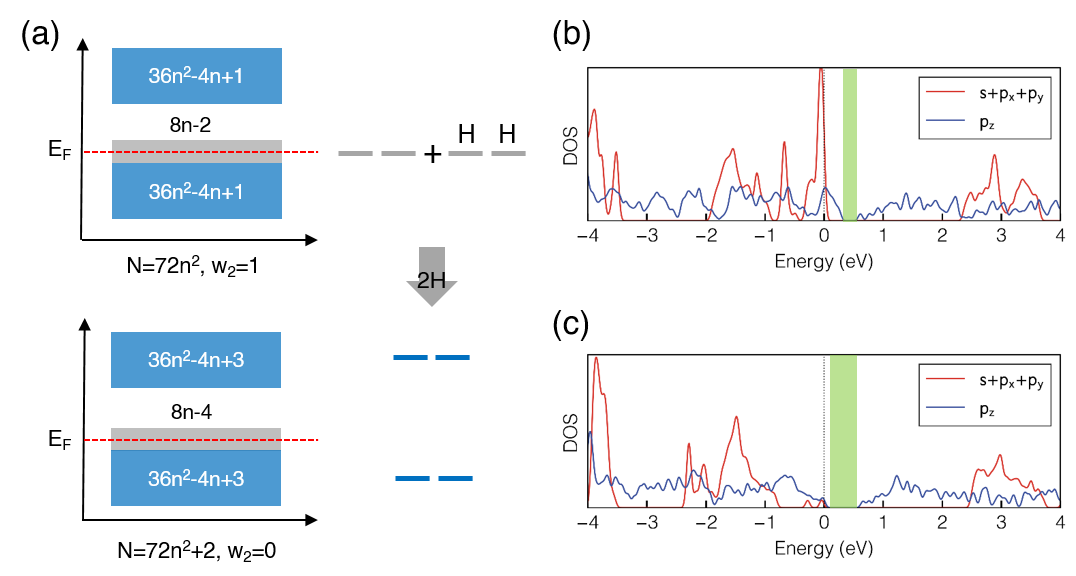}
\caption{
Filling anomaly and density of states.
(a) A schematic figure displaying the energy spectrum of a finite-size MGD composed of $n\times n$ unit cells, where the total number of states is $N=72n^2$. Here a thick gray strip near the Fermi level $E_F$ indicates the non-bonding states arising from carbon atoms along the boundary. The upper panel corresponds to the case at half-filling where an odd number of holes is below the gap, which demonstrates the higher-order band topology with $w_2=1$. The lower panel shows the case when two hydrogen atoms are added to the finite-size MGD. Here the total number of states $N'$ increases by two ($N'=N+2$). The hybridization between two hydrogen states and two non-bonding states generates two bonding and two anti-bonding states, so that the number of states below the gap changes from $\frac{N}{2}+4n-1$ to $\frac{N'}{2}+4n-2$. Then, $w_2$ also changes from 1 to 0. Thus, to maintain the value of $w_2$, the number of hydrogen atoms attached for passivation should be an integer multiple of four.
(b) Density of states (DOS) of a finite-size MGD without hydrogen passivation. Here the carbon atom at each corner has a non-bonding state. There are $(4n-1)$ holes below the gap at half-filling. The green color indicate the gapped region. (c) DOS of a finite-size MGD with hydrogen passivation where $(8n-4)$ hydrogen atoms are attached along the boundary. Here only two carbon atoms at $M_y$-invariant corners have non-bonding states.  There is a single hole below the gap at half-filling. Since both systems in (b) and (c) have an odd number of holes, both are HOTIs with $w_2=1$.
}\label{fig:DOS}
\end{figure*}

For the finite-size MGD composed of $n\times n$ unit cells shown in Fig.~\ref{fig:finitesize}(a), the total number of states are $N=72n^2$ when $2s$, $2p_x$, $2p_y$, and $2p_z$ orbitals are considered.
In MGD, there is a single $sp$ electron localized at every edge of a hexagonal unit cell. If an edge of a unit cell is not shared with an adjacent unit cell, the relevant $sp$ orbital becomes a half-filled zero-energy non-bonding state when chiral symmetry exists.
When we consider the geometry of a finite-size structure shown in Fig.~\ref{fig:finitesize}(a), MGD has $8n-2$ non-bonding states along the boundary.
When chiral symmetry is broken, $8n-2$ states can be merged into the valence band, so that $n_{\text{gap}}=36n^2+4n-1=\frac{N}{2}+4n-1$. This is what is observed in first-principles calculations and the corresponding density of states is shown in Fig.~\ref{fig:DOS}. Thus, when all the states below the gap are occupied, and thus the filling anomaly is lifted, a half-integral charge (modulo an integer charge) is accumulated at two $M_y$-invariant corners. This can be easily understood because an odd number of non-bonding states exists only at the $M_y$-invariant corners. [See Fig.~\ref{fig:finitesize}(a).]

To observe the corner states more clearly, we add a hydrogen atom to the carbon atom at every corner, except at two $M_y$-invariant corners, and keep $P$ and $M_y$ symmetries. Then the number of added hydrogen atoms is an integer multiple of four, which does not alter the band topology of MGD.
On the other hand, if two hydrogen atoms are added at the $M_y$-invariant corners, the total number of bands becomes $N'=N+2$, while $n_{\text{gap}}$ remains the same as described in Fig.~\ref{fig:DOS}(a). Then $n_{\text{gap}}=\frac{N'}{2}+4n-2$ and $w_2$ becomes trivial.
Therefore, the number of added hydrogen atoms should be an integer multiple of four to keep $w_2$ unchanged.
Maximally, $(8n-4)$ hydrogen atoms can be added to the finite-size MGD with two non-bonding states remaining at $M_y$-invariant corners.
The corresponding density of states is shown in Fig.~\ref{fig:DOS}(c). Here a half electric charge is accumulated at each $M_y$-invariant corner when the states below the gap are fully occupied as shown in Fig.~\ref{fig:finitesize}(b).

\subsection{Wannier function description}
Here we show that the higher-order band topology is associated with the chemical bonding of the $s$, $p_x$, and $p_y$ orbitals across the unit cell boundary.
In other words, Wannier functions are located away from the atomic sites and cannot be moved to the atomic sites by a continuous deformation that preserves inversion symmetry.
Here we follow the idea proposed by van Miert and Ortix~\cite{van2018higher} to analyze the Wannier centers in inversion-symmetric systems.

Let us take the inversion-invariant unit cell shown in Fig.~1(a) whose center is located at the position ${\bf m}$.
There are four Wyckoff positions in a unit cell labelled as $A, B, C,$ and $D$ ---located at ${\bf 0}$, $\frac{1}{2}{\bf a_1}$, $\frac{1}{2}{\bf a_2}$, and $\frac{1}{2}{\bf a_1}+\frac{1}{2}{\bf a_2}$ from the center of the unit cell, respectively --- which are invariant under inversion up to lattice vectors ${\bf a}_1$ and ${\bf a}_2$.
The symmetric Wannier functions centered at Wyckoff positions transform under inversion as follows:
\begin{align}
    w_{A\pm,{\bf m}}({\bf -x})&=\pm w_{A\pm,{\bf -m}}({\bf x}),\nonumber\\
    w_{B\pm,{\bf m}}({\bf -x})&=\pm w_{B\pm,{\bf -m-a_1}}({\bf x}),\nonumber\\
    w_{C\pm,{\bf m}}({\bf -x})&=\pm w_{C\pm,{\bf -m-a_1-a_2}}({\bf x}),\nonumber\\
    w_{D\pm,{\bf m}}({\bf -x})&=\pm w_{D\pm,{\bf -m-a_2}}({\bf x}),
\end{align}
where $w_{\alpha\pm,{\bf m}}({\bf x})$ denotes a Wannier function with the coordinate ${\bf x}$, centered at the Wyckoff position $\alpha$ of the unit cell with the center at ${\bf m}$.
The sign $\pm$ indicates eigenvalues of the inversion operation with respect to the Wyckoff position.
We define the number of symmetric Wannier functions centered at each Wyckoff position with inversion eigenvalues $\pm1$ as $N_{A\pm}, N_{B\pm}, N_{C\pm}$, and $N_{D\pm}$.
Let us note that two Wannier functions with inversion eigenvalues $+1$ and $-1$ at $A$ can be expressed as a linear combination of two Wannier functions with inversion eigenvalues $+1$ and $-1$ at any other Wyckoff position.
Thus, only the difference $\nu_{\alpha}=N_{\alpha+}-N_{\alpha-}$ at each Wyckoff position $\alpha$ is a well-defined $\mathbb{Z}$ invariant~\cite{van2018higher}.
\begin{align}
\nu_A=N_{A+}-N_{A-},\nonumber\\ \nu_B=N_{B+}-N_{B-},\nonumber\\ \nu_C=N_{C+}-N_{C-},\nonumber\\ 
\nu_D=N_{D+}-N_{D-},
\end{align}
which are related to the inversion eigenvalues at TRIMs in the Brillouin zone as~\cite{van2018higher}
\begin{align}
\nu_A
&=-\Gamma_1-\frac{1}{2}\Gamma_{-1}+\frac{1}{2}M_{-1}+\frac{1}{2}X_{-1}+\frac{1}{2}Y_{-1}=-1,\notag\\
\nu_B
&=\frac{1}{2}\Gamma_{-1}-\frac{1}{2}M_{-1}+\frac{1}{2}X_{-1}-\frac{1}{2}Y_{-1}=-1,\notag\\
\nu_C
&=\frac{1}{2}\Gamma_{-1}-\frac{1}{2}M_{-1}-\frac{1}{2}X_{-1}+\frac{1}{2}Y_{-1}=-1,\notag\\
\nu_D
&=\frac{1}{2}\Gamma_{-1}+\frac{1}{2}M_{-1}+\frac{1}{2}X_{-1}-\frac{1}{2}Y_{-1}=-1.
\end{align}
In monolayer graphdiyne (MGD), we find that $\nu_A=\nu_B=\nu_C=\nu_D=-1$ from the inversion eigenvalues at TRIM.
In the viewpoint of chemistry, $|\nu_A|=1$ comes from the electrons on the benzene ring around the unit cell center, $|\nu_B|=|\nu_C|=|\nu_D|=1$ originates from the $sp$ bonding across the unit cell boundary.

To relate these indices with the two-dimensional topological invariant $w_2$ defined before, we define electric dipole moments $p_{i=1,2}$ and an electric quadrupole moment $q_{12}$ for the unit cell as follows,
\begin{align}
    p_{x}&\equiv \sum_{\substack{i\in \rm occ,\\\rm unit\; cell}}X_i=\frac{1}{2}(N_B+N_C) =0\mod 1,\nonumber\\
    p_{y}&\equiv \sum_{\substack{i\in \rm occ,\\\rm unit\; cell}}Y_i=\frac{1}{2}(N_C+N_D) =0\mod 1,\nonumber\\
    q_{xy}&\equiv \sum_{\substack{i\in \rm occ,\\\rm unit\; cell}}X_iY_i=\frac{1}{4}N_C = \frac{1}{4}\mod \frac{1}{2},
\end{align}
where $X_i$ and $Y_i$ are the ${\bf a}_1$ and ${\bf a}_2$ components of the Wannier center for the $i$th occupied Wannier function.
$p_x$ and $p_y$ are nothing but the polarizations along ${\bf a}_1$ and ${\bf a}_2$, respectively.
$w_2$ is related to $p_{x,y}$, $q_{xy}$ via $w_2=4(p_xp_y-q_{xy})$ mod 2~\cite{ahn2018band,ahn2019failure}, so
\begin{align}
w_2
&=
(N_B+N_C)(N_C+N_D)-N_C\notag\\
&=
N_BN_C+N_CN_D+N_DN_B\mod 2\notag\\
&=
\nu_B\nu_C+\nu_C\nu_D+\nu_D\nu_B\mod 2,
\end{align}
because $\nu=N_+-N_-=N_++N_--2N_-=N-2N_-=N$ mod 2.
This shows that $w_2=1 \mod 2$ when $|\nu_B|=|\nu_C|=|\nu_D|=1$ due to the $sp$ bonding across the unit cell boundary.

\section{Discussion}

\begin{figure}[h]
\centering
\includegraphics[width=8.5 cm]{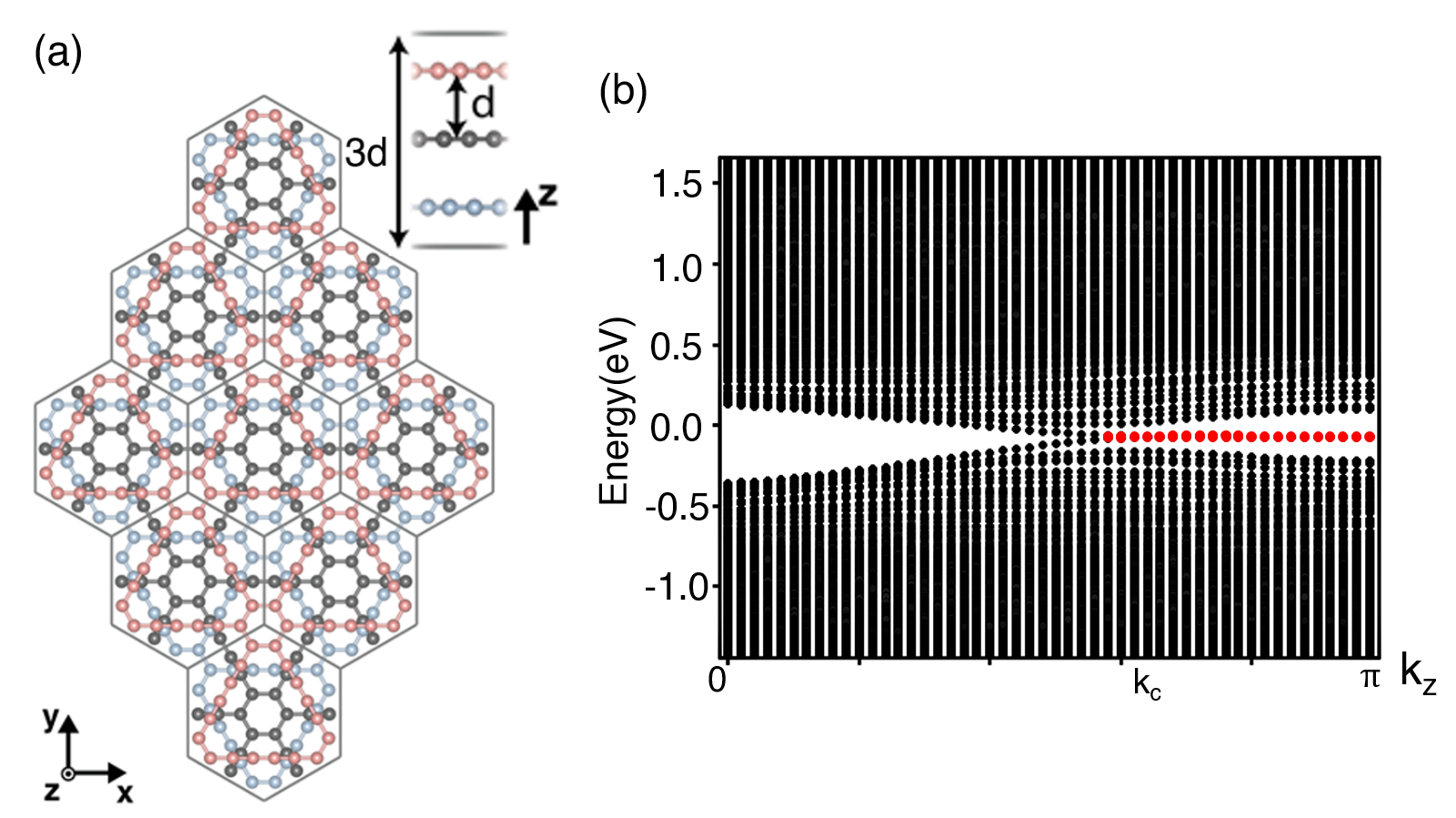}
\caption{
Hinge modes of ABC-stacked graphdiyne with monopole nodal lines.
(a) A schematic figure describing the 3D geometry of ABC-stacked graphdiyne which is finite in the $xy$ plane but periodic in the vertical direction. The structure must preserve $C_{2x}$ symmetry to carry hinge states. (b) Hinge modes, located in the region $k_c<|k_z|<\pi$, obtained by tight-binding Hamiltonian using only $p_z$ orbitals. In real materials including all core electronic levels, the hinge modes should appear in the region $0<|k_z|<k_c$ where $w_2=1$.
}\label{fig:ABCstack}
\end{figure}

Since MGD has $w_2=1$, when layers of MGD are stacked vertically, the resulting 3D insulator with $PT$ symmetry becomes a 3D weak topological insulator~\cite{ahn2018band}, dubbed a 3D weak Stiefel-Whitney insulator, when inter-layer coupling is weak. When inter-layer coupling becomes large enough, an accidental band crossing can happen at a TRIM at which a pair of nodal lines with $Z_{2}$ monopole charge are created. The existence of two nodal lines at $k_z=\pm k_c$ $(k_c>0)$ was predicted in ABC-stacked graphdiyne~\cite{nomura2018three}, and their $Z_2$ monopole charge was also confirmed~\cite{ahn2018band} based on first-principles calculations.
In ABC-stacked graphdiyne, the 2D subspace with a fixed $k_z$ carries $w_2=1$ ($w_2=0$) when $|k_z|<k_c$ ($k_c<|k_z|<\pi$) since the band inversion is happened at $\bm{k}=(0,0,\pi)$. Since a 2D HOTI with $w_2=1$ possesses corner charges, similar corner charges are also expected in ABC-stacked graphdiyne in the subspace with a fixed $k_z$ where $w_2=1$~\cite{wang2019higher}, which leads to hinge modes of the 3D structure shown in Fig.~\ref{fig:ABCstack}(a). 

To demonstrate the hinge modes of ABC-stacked graphdiyne, we study a tight-binding model by using only $p_z$ orbitals to reduce numerical costs. 
The energy spectrum of a finite-size system with $PT$ and $C_{2x}$ symmetries is shown in Fig.~\ref{fig:ABCstack}(b), which clearly shows the presence of hinge modes in the momentum region where $w_2=1$ as expected. When only $p_z$ orbitals are included, although the low-energy band structure from the tight-binding model is consistent with that from first-principles calculations, the topological property is different. Namely, the tight-binding model predicts $w_2=0$ ($w_2=1$) when $|k_z|<k_c$ ($k_c<|k_z|<\pi$), which is opposite to the result from the first-principles calculations. Such a discrepancy can be remedied when the core electronic levels are included in the tight-binding calculation, which is confirmed by separate calculations. 

To sum up, we have shown that the nontrivial band topology of a 2D HOTI lacking chiral symmetry can be revealed by using the filling anomaly, which can be explicitly demonstrated by counting the number of states of a finite-size system. This idea can be straightforwardly generalized to any $d$-th order topological insulator in $d$-dimensions hosting zero-dimensional corner states. We believe that our theoretical study provides a general way to extend the scope of HOTI materials to a wider class.

\section{Methods}
\subsection{DFT calculations}
We have carried out the first-principles density-functional theory calculations using the Vienna \textit{ab initio} simulation package~\cite{kresse1996efficient}.
The projector augmented-wave method~\cite{blochl1994projector} and the exchange-correlation functional of the generalized gradient approximation in the Perdew, Burke, and Ernzerhof~\cite{perdew1996generalized} scheme were used.
The self-consistent total energy was evaluated with an $12 \times 12 \times 1$ $k$-point mesh, and the cutoff energy for the plane-wave basis set was 500 eV.
For the finite-size structure, we used only $\Gamma$ point.
A single unit cell contains 18 carbon atoms, where a benzene ring is connected to six neighboring ones by linear chains.
In the optimized structure, we have
$L=9.46$ \AA,
$b_1=1.43$ \AA,
$b_2=1.40$ \AA,
$b_3=1.23$ \AA,
and $b_4=1.34$ \AA.
\\
\subsection{Tight-binding calculation}
Low-energy band structure of ABC-stacked graphdiyne can be properly described by $p_z$ orbital basis. 
The intralayer coupling is described by the nearest neighbor hopping parameters expressed as the transfer integral $V_{pp\pi}=V_{pp\pi}^0e^{(-R-a_0)/r_0}$, where $V_{pp\pi}^0\approx-2.7eV$ is the hopping parameter for graphene's nearest neighbor interaction ($a_0=0.142nm$ and $r_0=0.09nm$)~\cite{nomura2018three}.
The interlayer coupling is described by two hopping parameters of two shortest vertical bonds which are expressed as the transfer integral $V_{pp\sigma}=V_{pp\sigma}^0e^{(-R-d_0)/r_0}$, where $V_{pp\sigma}\approx0.48eV$  is the hopping parameter at the interlayer distance of graphite ($d_0=0.335nm$)~\cite{nomura2018three}.

\section{Notes}During the preparation of our manuscript, we have found a related work~\cite{sheng2019two}. In [\onlinecite{sheng2019two}], the authors also
observed HOTI phase in MGD. However, some of their theoretical arguments are inconsistent with our results. For instance, in [\onlinecite{sheng2019two}], it is stated that crystalline symmetries are not essential for the protection of corner states, which is in sharp contrast to our paper where the role of $P$ symmetry is emphasized.
Also, in [\onlinecite{sheng2019two}], it is mentioned that double band inversion of $p_z$ orbitals at the $\Gamma$ point is the origin of the HOTI phase. This
is different from our result in that higher-order band topology originates from the core levels derived from $s, p_x, p_y$ orbitals.

\section{Data Availability}
The datasets generated during and/or analyzed during the current study are available from the corresponding author on reasonable request.

\section{Acknowledgements}
We acknowledge the helpful discussions with Dongwook Kim and Youngkuk Kim.

\section{Competing Interests}
The Authors declare no competing interests.

\section{Author Contributions}
E.L. performed tight-binding calculations, R.K. performed first-principles calculations, and E.L. and J.A. did theoretical analysis. B.-J.Y. supervised the project. All authors analyzed the data and wrote the manuscript.\\
$^*$These authors contributed equally.

\section{Funding}
E. L., R. K. and J.A. were supported by IBS-R009-D1.
B.-J.Y. was supported by the Institute for Basic Science in Korea (Grant No. IBS-R009-D1) and Basic Science Research Program through the National Research Foundation of Korea (NRF) (Grant No.0426-20180011), and  the POSCO Science Fellowship of POSCO TJ Park Foundation (No.0426-20180002).
This work was supported in part by the U.S. Army Research Office under Grant Number W911NF-18-1-0137.


\end{document}